

\magnification\magstep1
\hsize 27pc
\vsize 43pc
\nopagenumbers
\headline{\ifodd\pageno\rightheadline \else\leftheadline\fi}
\footline{\hfil}

\voffset .2 truein
\hoffset .55truein

\font\fb=cmr12
\font\fc=cmr10
\font\fab=cmr9
\font\fs=cmr6
\font\fd=cmr5
\font\bfab= cmbx9

\def\eightpoint{\def\rm{\fam0\eightrm}
  \font\eightrm=cmr9
  \font\sixrm=cmr6
  \font\fiverm=cmr5
  \font\eighti=cmmi9
  \font\sixi=cmmi6
  \font\fivei=cmmi5
  \font\eightsy=cmsy9
  \font\sixsy=cmsy6
  \font\fivesy=cmsy5
  \font\eightbf=cmbx9
		\font\sixbf=cmbx6
  \font\fivebf=cmbx5
  \font\eightit=cmti9
  \font\eighttt=cmtt9
  \font\eightsl=cmsl9
  \font\tentex=cmtex10
  \font\sevenrm=cmr7
  \textfont0 = \eightrm
  \scriptfont0 = \sixrm
  \scriptscriptfont0 = \fiverm
  \textfont1 = \eighti
  \scriptfont1 = \sixi
  \scriptscriptfont1 = \fivei
  \textfont2 = \eightsy
  \scriptfont2 = \sixsy
  \scriptscriptfont2 = \fivesy
  \textfont3=\tenex
  \scriptfont3=\tenex
  \scriptscriptfont3=\tenex
  \textfont\itfam=\eightit  \def\it{\fam\itfam\eightit}%
  \textfont\slfam=\eightsl  \def\sl{\fam\slfam\eightsl}%
  \textfont\ttfam=\eighttt  \def\tt{\fam\ttfam\eighttt}%
  \textfont\bffam=\eightbf  \scriptfont\bffam=\sixbf
  \scriptscriptfont\bffam=\fivebf  \def\bf{\fam\bffam\eightbf}%
  \normalbaselineskip=11pt
  \setbox\strutbox=\hbox{\vrule height8pt depth3pt width0pt}%
  \let\sc=\sevenrm  \let\big=\eightbig \normalbaselines\rm}

 \def\eightbig#1{{\hbox{$\textfont0=\tenrm\textfont2=\tensy
   \left#1\vbox to7.25pt{}\right.\n@space$}}}

\def\ninepoint{\def\rm{\fam0\ninerm}
  \font\ninerm=cmr9
  \font\sixrm=cmr6
  \font\fiverm=cmr5
  \font\ninei=cmmi9
  \font\sixi=cmmi6
  \font\fivei=cmmi5
  \font\ninesy=cmsy9
  \font\sixsy=cmsy6
  \font\fivesy=cmsy5
  \font\ninebf=cmbx9
		\font\sixbf=cmbx6
  \font\fivebf=cmbx5
  \font\nineit=cmti9
  \font\ninett=cmtt9
  \font\ninesl=cmsl9
  \font\tentex=cmtex10
  \font\sevenrm=cmr7
  \textfont0 = \ninerm
  \scriptfont0 = \sixrm
  \scriptscriptfont0 = \fiverm
  \textfont1 = \ninei
  \scriptfont1 = \sixi
  \scriptscriptfont1 = \fivei
  \textfont2 = \ninesy
  \scriptfont2 = \sixsy
  \scriptscriptfont2 = \fivesy
  \textfont3=\tenex
  \scriptfont3=\tenex
  \scriptscriptfont3=\tenex
  \textfont\itfam=\nineit  \def\it{\fam\itfam\nineit}%
  \textfont\slfam=\ninesl  \def\sl{\fam\slfam\ninesl}%
  \textfont\ttfam=\ninett  \def\tt{\fam\ttfam\ninett}%
  \textfont\bffam=\ninebf  \scriptfont\bffam=\sixbf
  \scriptscriptfont\bffam=\fivebf  \def\bf{\fam\bffam\ninebf}%
  \normalbaselineskip=11pt
  \setbox\strutbox=\hbox{\vrule height8pt depth3pt width0pt}%
  \let\sc=\sevenrm  \let\big=\ninebig \normalbaselines\rm}

 \def\ninebig#1{{\hbox{$\textfont0=\tenrm\textfont2=\tensy
   \left#1\vbox to7.25pt{}\right.\n@space$}}}


\def\abs#1{\bigskip\bigskip\medskip{\ninepoint%
       {\narrower \noindent%
      {\bf Abstract.} #1 \medskip\bigskip}}\medskip}

\def\Acknowledgments#1{\goodbreak\bigskip\noindent{%
{\bf Acknowledgments.} #1.}}  


\def\ats{\kern.15em\hbox{@}\kern.15em}

\def\aut{\bigskip \bigskip
        \centerline {\author} }

\def\CC{\hbox{\rm C\kern -.58em {\raise .54ex \hbox{$\scriptscriptstyle |$}}
  \kern-.55em {\raise .53ex \hbox{$\scriptscriptstyle |$}} }}

\def\copyright{\hbox{{\fb o}\kern-.61em \raise .46ex \hbox{\fd c}}}

\def\cor#1 #2{\begingroup\medbreak\noindent{\bfab Corollary #1 }\sl #2}

\def\defin#1 #2{\begingroup\medbreak\noindent{\bfab Definition #1 } #2}

\rm

\rm



\def\footnoterule{\kern -3pt \hrule width 0truein \kern 2.6pt}

\def\frac#1#2{{#1 \over #2}}

\def\lem#1 #2{\begingroup\medbreak\noindent{\bfab Lemma #1 }\sl #2}

\def\mbf#1{\setbox0=\hbox{#1}%
           \kern-.025em\copy0\kern-\wd0
           \kern.05em\copy0\kern-\wd0
           \kern-.025em\hbox{\raise.0433em\box0} }

\def\NN{{{\rm l}\kern-.15em{\rm N}}}

\def\nullx{\hfill}

\def\per{\kern .03em \% \kern .02em}





\def\pro#1 #2{\begingroup\medbreak\noindent{\bfab Proposition #1 }\sl #2}

\def\References{\goodbreak\bigskip\centerline{\bf References}\medskip
   \frenchspacing \hfuzz 2pt}

\rm

\def\rightheadline{\ifnum\pageno=\count100 \nullx%
  \else\it\chptitle\hfil\rm\folio\fi}
  \def\leftheadline{\ifnum\pageno=\count100 \nullx%
  \else\rm\folio\hfil\it\authead\fi}  

\def\RR{{{\rm l}\kern-.17em{\rm R}}}

\def\sRR{{\sl \hbox{I\kern-.2em\hbox{R}}}}

\def\sect#1{\goodbreak\bigskip\smallskip\centerline{\bf\S
#1}\bigskip\noindent\ignorespaces}


\def\tabone#1 #2{\goodbreak\medskip\centerline{\bfab Table #1.}\smallskip
  \centerline {#2.} \smallskip}

\def\tabtwo#1 #2 #3{\goodbreak\medskip\centerline{\bfab Table #1.}\smallskip
   \centerline {#2} \vskip.1pt
    \centerline {#3.}\smallskip}

\def\tabthre#1 #2 #3 #4{\goodbreak\medskip\centerline{\bfab Table
   #1.}\smallskip
   \centerline {#2} \vskip.1pt
   \centerline {#3} \vskip.1pt
   \centerline {#4.}\smallskip}

\def\theo#1 #2{\begingroup\medbreak\noindent{\bfab Theorem #1 }\sl #2}
 \def\tit#1{\centerline {\bf #1}}
\def\titwo#1{\medskip \centerline {\bf #1}}

\def\titexp#1#2{\hbox{{\bf #1} \kern-.25em \raise .90ex \hbox{\bf #2}}\/}
\def\titsub#1#2{\hbox{{\bf #1} \kern-.25em \lower .60ex \hbox{\bf #2}}\/}

\def\ZZ{{{\rm Z}\kern-.52em{\rm Z}}}




\centerline{}
\pageno=\count100
\count102=\count100
\advance\count102 by \count101
\advance\count102 by -1
\insert\footins{\fs
\medskip
\baselineskip 8pt
\leftline{{Approximation Theory VIII}
\hfill {\fc \the\pageno}}
\leftline{Charles K. Chui and Larry L. Schumaker\ (eds.),
    pp. \the\pageno--\the\count102.}
\leftline{Copyright \copyright\ 1995 by World Scientific Publishing Co., Inc.}
\leftline{All rights of reproduction in any form reserved.}
\leftline{ISBN 0-12-xxxxxx-x}
\smallskip
\par\allowbreak}

\def\ref{\global\advance\refnum by 1 \item{\the\refnum .}}
\newcount\refnum \refnum = 0

\def\em{\it}

\def\eopp{{ \vrule height7pt width7pt depth0pt} }

\def\xbf{{\bf x}\,}

\def\fh{\hat{f}}
\def\fhl{\widehat{f^{(\ell)}}}

\def\part{\partial}
\def\kh{\hat{k}}

\tit{Piecewise Convex Function Estimation}
\titwo{and Model Selection}
\def\chptitle{Piecewise Convex Model Selection} 
\def\author{Kurt S.~Riedel}
\def\authead{Kurt S.~Riedel}
{\vskip-.12in}
\aut
{\vskip-.26in}
\abs{Given noisy data, function estimation is considered when the unknown
function is known {\it a priori} to consist of a small number of regions
where the function is either convex or concave. When the regions are known
{\it a priori}, the estimate is reduced to a finite dimensional convex
optimization in the dual space. When the number of regions is unknown, the
model selection problem is to determine the number of convexity change
points. We use a pilot estimator based on the 
expected number of false inflection points.
}
{\vskip-.40in}
\sect{1. Introduction}
{\vskip-.1in}

Our basic tenet is:
``Most real world functions are piecewise $\ell$-convex with a small number of
change points of convexity.''
Given $N$ measurements of the unknown function, $f(t)$, contaminated with
random noise, we seek to estimate $f(t)$ while preserving the geometric
fidelity of the estimate, $\fh(t)$, with respect to the true function. In
other words, the number and location of the change points of convexity of
$\hat{f}(t)$ should approximate those of ${f}(t)$.

We say
that $f(t) $ has $k$ change points of $\ell$-convexity  with change points
$x_1 \leq x_2 \ldots \leq x_k$ if $(-1)^{k-1} f^{(\ell )} (t) \geq 0$ for
$x_k \leq t \leq x_{k+1}$.
For $\ell = 0$, $f(t)$ is
nonnegative and for $\ell =1$, the function is nondecreasing. In regions
where the constraint of $\ell$-convexity is active, $f^{(\ell )} (t) = 0$ and
$f(t)$ is a polynomial of degree $\ell -1$. For 1-convexity, $f(t)$ is constant in
the active constraint regions and for 2-convexity, the function is linear.
Our subjective belief is that most people prefer smoothly varying functions
such as quadratic or cubic polynomials even in the active constraint regions.
Thus, piecewise 3-convexity or 4-convexity are also reasonable hypotheses.
The idea of constraining the function fit to preserve $\ell$-convexity
properties has been considered by a number of authors. The more
difficult problems of determining the number and location of the
$\ell$-convexity breakpoints will be a focus of this article. We refer to
the estimation of the number of change points as the
``model selection problem'' because it resembles model selection in an
infinite family of parametric models.
{\vskip-.1in}
\sect{2 Convex Analysis}
{\vskip-.1in} 
In this section, we assume that the change points $\{ x_1 \ldots x_k \}$
of $\ell$-convexity are given and that the function is in the Sobolev space,
$W_{m,p} [0,1]$ with $m\geq \ell$ and $1<p< \infty$ where
{\vskip-.22in} 
$$
W_{m,p} = \{  f|f^{(m)}  \in L_p [0,1] \ {\rm and\ } f,f^{\prime} \ldots
f^{(m-1)} (t) \ {\rm  \ absolutely \ continuous} \} \ .
$$
{\vskip-.1in} 
\noindent
We decompose $W_{m ,p}$
into a direct sum of the space of polynomials of degree $m-1$, $P_{m-1}$
plus the set of functions whose first $m-1$ derivatives vanish at $t=0$
which we denote by $W_{m,p}^0$ [10]. 

Given change points, $\{ x_1 ,x_2 \ldots x_k \}$, we define the closed
convex cone
{\vskip-.2in} 
$$
V^{k,\ell }_{m,p} [x_1 ,\ldots ,x_k ] = \{ f \in
W_{m,p}  \ | \ (-1)^{k-1}
f^{(\ell )} (t) \geq 0 \ \ {\rm for } \ \ x_{k-1} \leq t < x_k \} \ .
$$
{\vskip-.1in}
Let $\bf{x}$  denote the $k$ row vector, $( x_1 ,x_2 \ldots x_k )$.
We define the class of functions with at most $k$ change points as
{\vskip-.2in} 
$$
V_{m,p}^{k,\ell} \equiv
\bigcup_{x_1 \leq x_2 \ldots \leq x_k }
\bigl\{ V^{k,\ell}_{m,p} [x_1 ,\ldots ,x_k ] \cup (
-V^{k,\ell}_{m,p} [x_1 ,\ldots ,x_k ] ) \bigr\} \ .
$$
{\vskip-.15in} 
By allowing $x_{k^{\prime} }
= x_{k^{\prime} +1} $, we have embedded
$V^{k,\ell}_{m,p}$ into
$V^{k+1,\ell}_{m,p}$.
$V^{k,\ell}_{m,p}$
is the union of convex cones, and is closed but not convex. 
For the case $p = \infty$, similar piecewise $\ell$-convex classes are
defined in [2].
To  decompose $W_{m,p}$ in terms of $V^{k,\ell}_{m,p}$,
we require that each function in $W_{m,p}$
has a piecewise continuous $\ell$-th derivative.
By the Sobolev embedding theorem,
this corresponds to the case $m\geq \ell +1$.

Let $\|f\|^p_{j,p} \equiv \int_0^1 |f^{(j)} (t) |^p dt$.
We endow $W_{m,p}$ with the  norm:
{\vskip-.15in} 
$$ 
|\| f \|\vert^p_{m,p} = \sum_{j=0}^{m-1} |f^{(j)} (t=0) |^p +
\|f\|^p_{m,p}         
\ .
$$
{\vskip-.1in}
\noindent 
The dual space of $W_{m,p}$ is isomorphic to the direct sum 
of $P_{m-1}$ and $W_{m,q}^0$ with $q=p/(p-1)$ and the duality pairing: 
{\vskip-.15in}
$$\langle\langle g,f\rangle\rangle = 
\sum_{j=0}^{m-1} b_j a_j + \int_0^1 f^{(m)} (t) g^{(m)} (t) dt 
\ .  \eqno (2.1)
$$
{\vskip-.1in}\noindent
In (2.1), $f\in W_{m,p}$,  $g \in W_{m,q}$, $a_j \equiv f^{(j)}(0)$ 
and $b_j \equiv g^{(j)}(0)$. 
We denote the
duality pairing by $\langle\langle \cdot\rangle\rangle $  and the $L_2$ inner
product by $\langle \cdot\rangle $.
The space
$W_{m ,p}$
has a reproducing kernel, $R(t,s)$, such that for each $t$, $f(t) =
\langle \langle R_t ,f\rangle\rangle$ [10].
A linear operator, $L_i^* \in
W_{m,p}^*$ has representations 
$L_i f = \langle\langle L_i R(\cdot ,s),f(s)\rangle\rangle$ and
$L_i f = \langle L_i \delta( s- \cdot),f(s)\rangle$. 

We are given $n$ measurements of $f(t)$:
{\vskip-.1in} 
$$
 \ y_i = L_if  + \epsilon_i =
\langle h_i ,f \rangle + \epsilon_i  =
\langle\langle m_i ,f \rangle\rangle + \epsilon_i \ , \eqno(2.2)
$$
{\vskip-.1in} \noindent
where $L_i R(\cdot ,s)$ are linear operators 
in $W_{m,p}^{\perp}$, 
and the $\epsilon_i$ are independent, normally distributed random
variables with variance $\sigma_i^2 > 0$.
We represent $L_i$ as
 $m_i(s) \equiv  L_i R(\cdot ,s)$ and  
$h_i(s) \equiv  L_i \delta (s-\cdot )$ and assume 
$h_i \in W_{\ell,1}^{\perp}$. In the standard case where
$y_i = f(t_i ) + \epsilon_i$, $m_i(s) = R(t_i,s)$ and $h_i(s) =\delta(s-t_i)$. 

\noindent
A robustified estimate of $f(t)$ given the measurements $\{y_i\}$ 
is $\hat{f} \equiv{\rm argmin}$ ${\rm{VP}}
[f\in V^{k,\ell}_{m,p} [x_1 ,\ldots ,x_k ] ]$:
{\vskip-.15in} 
$$
{\rm VP}[f ] \equiv
\frac{\lambda}{p}\int |f^{(m)} (s) |^p ds +
 \sum_{i=1}^N \psi_i \left( \langle h_i ,f \rangle -y_i \right) 
\ , \ \
\eqno(2.3)$$
{\vskip-.12in} 
\noindent
where the $\psi_i$ are strictly convex, continuous functions. The
standard case is $p=2$ and $\psi_i (y_i -\langle h_i ,f\rangle ) =
|y_i -f(t_i )|^2 /n\sigma_i^2$.
The set of $\{ h_i , i=1,\ldots ,N \}$ separate polynomials of degree
$m-1$ means that  $\langle h_i , \sum_{k=0}^{m-1} c_k t^k \rangle
= 0$, $\forall i$ implies $c_k \equiv 0$. 

\noindent
{\bf Theorem 1.}{\em Let $\{ h_i \}$ separate
polynomials of degree $m-1$, then the {mini}mization problem ({2.3}) has
an unique solution in $V^{k,\ell}_{m,p} [\xbf]$ and  the
minimizing function is in $C^{2m-\ell-2}$ and
satisfies the differential equation:
{\vskip-.192in} 
$$   
(-1)^m d^m [|\fh^{(m)} |^{p-2} \fh^{(m)} (t) ] + \sum_{i=1}^n \psi_i^{\prime}
(\langle h_i ,\fh \rangle -y_i ) h_i(t) = 0 \ ,
\eqno (2.4)$$
{\vskip-.15in} \noindent
in those regions where $|f^{(\ell )} |>0$ for  $1 < p < \infty$.}

\noindent
{\bfab Proof:} The functional ({2.3}) is strictly convex, lower semicontinuous
and coercive, so by Theorem 2.1.2 of Ekeland and Temam, it has a unique
minimum, $f_0$, on any closed convex set. From the generalized calculus
of convex analysis, the solution satisfies
{\vskip-.18in} 
$$
0 \in (-1)^m d^m [(|f^{(m)} |^{p-2} f^{(m)} (t)] 
\Sigma \psi_i^{\prime} (\langle h_i ,\fh \rangle  -y_i ) h_i(t) +
\partial N_V(f) 
\eqno(2.4)
$$
{\vskip-.1in} \noindent
where $N_V(f)$ is the normal cone of  $V^{k,\ell}_{m,p} [\xbf]$ at $f$
[1, p.~189].    
From [9], each element of $N_V(f)$ is the limit of a discrete sum:
$\sum_{t} a_t \delta^{(\ell)}(\cdot-t)$ where the $t's$ are in the active
constraint region. Integrating (2.4) yields
{\vskip-.18in} 
$$\eqalignno{ |f^{(m)} |^{p-2} f^{(m)} (t)  &=
\sum_{i=1}^n { \psi_i^{\prime} (\langle h_i ,\fh \rangle  -y_i ) 
\langle h_i(s), (s-t)_+^{m-1}\rangle \over (m-1)!} \cr &
+ \int {(s-t)_+^{m-\ell-1}d\mu(s) \over  (m-\ell-1)!} \  ,
& (2.5)}
$$
{\vskip-.1in} \noindent
where $d\mu$ corresponds to a particular element of  $N_V(f)$.
Since $ (s-t)_+^{m-\ell-1}$ is $m-\ell-2$ times differentiable, 
the right hand side of (2.5) is $m-\ell-2$ times differentiable.
Integrating (2.5) yields $f\in C^{2m-\ell-2}$.
\eopp


The intervals on which $f^{(\ell )} (t)$ vanishes are unknowns and need
to be found as part of the optimization. Using the differential
characterization ({2.3}) loses the convexity properties of the underlying
functional. For this reason, extremizing the dual functional is now
preferred.

\noindent
{\bf Theorem 2.} 
{\em 
The dual variational problem is: Minimize over $\alpha \in \RR^n$
{\vskip-.13in} 
$$
{\rm VP}^*[\alpha;{\bf x}] \equiv 
{\lambda^{1-q} \over q} \int |[{\bf P_x}^*M\alpha ]^{(m)} (s) |^q ds +
\sum_{i=1}^n \psi_i^* (\alpha_i ) - \alpha_i y_i \ ,
\eqno (2.6)$$
{\vskip-.1in} 
\noindent
where $M \alpha (t) \equiv \sum_i m_i (t) \alpha_i$ 
and $\psi^*_i$ is the
Fenchel/Legendre transform of $\psi_i$. The dual projection ${\bf P_x}^*$ is 
defined as 
{\vskip-.1in} 
$$  
\int |[{\bf P_x}^* g]^{(m)} (s) |^q ds 
\equiv {\rm inf}_{\tilde{g}\in V^-}  \int_0^1 |g^{(m)}
-\tilde{g}^{(m)} (s)|^q 
\ ,
\eqno(2.7)$$
{\vskip-.1in} \noindent 
where the minimization is over $\tilde{g}$ in the dual cone subject to 
$g^{(j)}(0)=\tilde{g}^{(j)}(0),\ \ 0\le j <m$.
The dual problem is strictly convex
and its minimum is the negative of the infimum of ({2.3}).
} 
{\vskip .05in} 

\noindent
{\bfab Proof:} Let $\psi_V$ be  the indicator function of $V^{k,\ell}_{m,p} [\xbf]$ 
and define
{\vskip-.11in} 
$$
U(f) = {\lambda \over p} \int_0^1 |f^{(m)} (s)|^p ds + \psi_V (f) \ .
\eqno(2.8)$$
{\vskip-.07in} \noindent 
We claim that the Legendre transform of $U(f)$ is the first term in (2.6).
Note that $\psi_V^* (g) = \psi_{V^-} (g)$, the indicator function
of the dual cone $V^-$. Since the Legendre transform of the first term in (2.8)
is
{\vskip-.15in} $$
V_1^* (g) = {\lambda^{1-q}\over q} \int_0^1 |g^{(m)} (s) |^q ds \ \ 
{\rm for}\ \ g \in W_{m,q}^0, \ \ \ {\rm and}\ \ \infty \ \ {\rm otherwise}.
$${\vskip-.1in} \noindent
Our claim follows from 
$[U_1 +U_2 ]^* (g) = \inf_{g'} \{ U_1^* (g-g' ) + U_2^* (g' ) \}$.
\noindent
The remainder of the theorem 
follows from the general duality theorem of
Aubin and Ekeland [1, 
p.~221]. 
\eopp

For the case $\ell=m$, the minimization over the dual cone can be done
explicitly. 
For $\ell<m$, 
Theorem 1 is proven in [9] and Theorem 2 is proven in [6]
for the case $p=2$ and $\psi (y) = y^2$.
Equation (2.5) and the corresponding smoothness results appear in [9]
for the case $\ell=1$, $p=2$ and $L_i = \delta(t-t_i)$.

{\vskip-.12in} 
\sect{3. Change point estimation}
{\vskip-.12in} 
When the number of change points is fixed, but the locations are unknown, we
can estimate them by minimizing the functional in ({2.3}) with respect to the
change point locations. We now show that there exists a set of minimizing
change points.

\noindent
{\bf Theorem 3.} {\em For each $k$, there exist change points 
$\{ x_{j},\ j =1,
\ldots k\}$ that minimize the variational problem ({2.3}).}

\noindent
{\bfab Proof:} We use the dual variational problem ({2.5}) 
and maximize over $\xbf \in [0,1]^k$ after minimizing over the 
$\alpha \in R^N$. 
The functional ({2.5}) is
jointly continuous in $\alpha ,\xbf$  and convex in $\alpha$. Theorem 3
follows from the min-max theorem [1,p. 296].
\eopp

The change point locations need not be unique. The proof requires $\le$ 
instead of $<$ in the ordering
$x_j \leq x_{j+1}$ to make the change point space compact. When
$x_j = x_{j+1}$, the number of effective change points is less than $k$.
Finding the $\xbf$ that minimizes VP$^*$ is computationally intensive
and requires the solution of a convex programming problem at each step.
Theorems 1-3 
are valid when $\ell \leq m$ including $\ell  =m$. 
Restricting to $p=2$, we have the following theorem from [9]:

\noindent
{\bf Theorem 4.} [Utreras] 
{\em Let $f$ be in a closed convex cone,  $V \subseteq W_{m,2}$,
let $\fh_u$ be the unconstrained minimizer of (2.3) given $y_i$
and $\fh_c$ be the constrained minimizer (with $p=2$ and $\psi_i(y)=
|y|^2/\sigma_i^2$). Then $\|f - \fh_c \|_V  \le \|f - \fh_u \|_V$
where $\|f  \|_V^2 \equiv
\frac{\lambda}{2}\int |f^{(m)} (s) |^2 ds +$ 
$ \sum_{i=1}^N \psi_i ( L_i f  ) $. 
}

Theorem 4 shows that if one is certain that $f$ is in a particular 
closed convex cone, the constrained estimate is always better than 
the unconstrained one. Unfortunately Theorem 4 does not generalize
to unions of convex cones and thus does not apply to $V^{k,\ell}_{m,2}$.

\sect{4. Number of false inflection points}

We now consider unconstrained estimates of $f(t)$ and examine the number 
of false $\ell$-inflection points. We assume that the noisy measurements
of $f$ occur at nearly regularly spaced locations, $t_i$
(with $h_i(t) = \delta(t-t_i))$. Specifically, we assume that 
$d_n \equiv sup_t \{ F_n(t) -F(t) \}$ tends to zero as $n^{-b}$ with
$b \ge 0$ where $F_n(t)$ is the empirical distribution of the $t_i$
and $F(t)$ is the limiting distribution. For regularly spaced points,
$d_n \sim 1/n$. This nearly regularly spaced assumption allows us to
approximate the discrete sums over the $t_i$ by integrals.

A smoothing kernel estimate of $f^{(\ell)}(t)$ is a 
weighted average of the $y_i$: 
{\vskip-.1in}
$$\fhl(t) = \sum_i y_i \kappa({t-t_i \over h_n})
\left[\frac{t_{i+1}-t_{i-1}}{2}\right] \ , \eqno(4.1)$$
{\vskip-.1in} \noindent
where $h_n$ is the kernel halfwidth and $\kappa$ is the kernel. 
$\kappa$ is required to satisfy the moment conditions:
$\int_{-1}^1 s^j \kappa(s)ds = \ell! \delta_{j,\ell}$, $0\le j < \ell +2$,
with $\kappa \in C^{2}[-1,1]$ and that 
$\kappa(\pm 1) =\kappa'(\pm 1) = 0$. We call such functions- $C^2$ 
extended kernels.
When $f\in C^{m}$, the optimal halfwidth scales as $h_n \sim n^{-1/(2m+1)}$, 
and the optimal spline smoothing parameter scales as 
$\lambda_n \sim n^{-2m/(2m+1)}$. 
In [5], Mammen et al.~derive the  number of false inflection points 
for kernel estimation of a probability density. We present the analogous
result for regression function estimation. The proofs in our case
are easier because we need only show that discrete sums converge to
their limits. Our results are for arbitrary $\ell$ while [4,5] considered
$\ell=1,2$.   
  
 {\bf Theorem 5.} (Analog of [4,5]) 
{\em Let $f(t) \in C^{\ell+1}[a,b]$ have $K$\ $\ell$-inflection
points $\{x_1, \ldots x_k\}$ with $f^{(\ell +1)}(x_j) \ne 0 $,
$f^{(\ell )}(a) \ne 0 $ and $f^{(\ell)}(b) \ne 0 $.
Consider a sequence of kernel smoother estimates with $C^2$ extended kernels.
Let the sequence of kernel halfwidths, $h_n$, satisfy
$0<{\rm liminf}_n h_n n^{1/(2\ell+3)} $
$\le{ \rm limsup}_n h_n n^{1/(2\ell+3)}< \infty $,
then the expected number of $\ell$-inflection points is 
$$ {\bf E}[\hat{K}] - K
=  \ 2\sum_{j=1}^K 
H\left(\sqrt{ {nh^{2\ell+3} | f^{(\ell +1)}(x_j)|^2
\over \sigma^2\|\kappa^{(\ell +1)}\| F'(x_j) } }
\right)  \ ,\eqno(4.2) $$
where $\sigma^2= {\bf Var}[\epsilon_i]$, 
$H(z) \equiv \phi(z)/z +\Phi(z) -1$ with $\phi$ and $\Phi$ being the 
Gaussian density and distribution provided that $d_n~< n^{-1/2}$.
}

\noindent
{\bfab Proof:} The proof consists of applying the Cram\'er-Leadbetter
zero-crossing formula to (4.1) and then taking the limit as 
$n \rightarrow \infty$.

\noindent
{\bf Theorem} (Cram\'er-Leadbetter) {\em Let $N$ be the number of zero crossings
of a differentiable Gaussian  process, $Z(t)$, in the time interval [0,T].
Then 
$$ {\bf E}[N] = \int_0^T \frac{\gamma(s)\rho(s)}{\sigma(s)} 
\phi\left(\frac{m(s)}{\sigma(s)}\right) G(\eta(s))ds  \ , \eqno(4.3)$$  
where
$\sigma^2(s) ={\bf Var}[Z(s)]$, $\gamma^2(s) ={\bf Var}[Z'(s)]$,
$\mu(s) ={\bf Corr}[Z(s)Z'(s)]$, $\rho(s)^2 = 1 - \mu(s)^2$,
$m(s) ={\bf E}[Z(s)]$, $\eta(s) = 
\frac{m'(s) - \gamma(s)\mu(s)m(s)/\sigma(s)}{\gamma(s)\rho(s)}$.   
}

We claim that for (4.1), 
$\sigma_n^2(s)  \rightarrow \sigma^2\|\kappa^{(\ell)} \|^2  F'(s)
/ n h^{2\ell +1}$, 
$\gamma_n^2(s) \rightarrow  \sigma^2$
$\|\kappa^{(\ell+1)} \|^2$ $F'(s) / n h^{2\ell +3} $,
$\mu_n(s) \rightarrow {\cal O}_R(h_n + 1/nh^{2\ell+1})$, 
$\rho_n(s)^2 \rightarrow 1$,
$m_n(s) \rightarrow f^{(\ell)}(s) + {\cal O}(h_n + 1/nh^{\ell+1}) $. 
To show the convergence of the discrete sums to integrals,
we use $\int g(s) ds = \sum_i g(t_i) [ t_{i+1} -t_{i-1}]/2
+ {\cal O}_R(sup_i [ t_{i+1} -t_{i-1}]^2/h_n^2)$ 
where ${\cal O}_R$ denotes a relative size of ${\cal O}$.
More detailed proofs of the convergence of 
the discrete sums to integrals can be found in [2]. 
Since the integrand in (4.3) is bounded and converging pointwise,
the dominated convergence theorem shows that 
the sequence of integrals given by (4.3) converges.  {Equation} (4.2)
follows by evaluating the integral using the method of steepest descent.
\eopp

{\bf Corollary.} 
{\em Let $f(t) \in C^{\ell+1}[a,b]$, $d_n/h_n^{\ell+1} \rightarrow 0$ 
and $nh_n^{2\ell+3} \rightarrow \infty$
with $\kappa$ a $C^2$ extended kernel. The probability that 
$\widehat{f^{(\ell +1)}}$ has a false inflection point outside of a width of 
$\delta$ from the actual $(\ell+1)$-inflection points is 
${\cal O}(\exp(-nh_n^{2\ell +3}))$. 
}

For the case $p=2$,
the smoothing spline estimate is a linear estimate of the form
$\fhl(t) = \sum_i y_i g_{n,\lambda}(t,t_i)$ where $ g_{n,\lambda}(t,t_i)$
solves the equation: 
$ (-1)^m \lambda_n g_{n,\lambda}^{(2m)}(t,s) +  
\sum_i^n g_{n,\lambda}(t_i,s) = \delta(t-s)$,
with the boundary conditions, $\partial_t^j g_{n,\lambda}(0,s)=
0 =\partial_t^j g_{n,\lambda}(1,s)$ for $m \le j <2m$.

{\bf Theorem 6.} [Silverman]
{\em  Let $\lambda_n^{-1/2m} d_n \rightarrow 0$, $|F''(t)|<\infty$  
and $0<c_1 < F'(t) < c_2 $. 
The Green's function, $ g_{n,\lambda}(t,t_i)$, of the smoothing spline 
converges 
to a kernel function with
the  halfwidth, $h(t) =[\lambda F'(t)]^{1/2m}$: 
{\vskip-.06in} 
$$ 
h(t)\ \partial_s^{j} g_{n,\lambda}(t+h(t)s,t) 
\rightarrow \partial_s^{j}\kappa(s)/ F'(t)\ , \ \ \ 0 \le j <m$$
{\vskip-.05in} \noindent
where
the equivalent  {kernel} satisfies 
$(-1)^m\kappa^{(2m)}(t) + \kappa(t) = \delta(t)$ with decay at infinity 
boundary conditions. The convergence is uniform for
in any closed subdomain, $t \in [\delta,1-\delta]$ and 
$t+h(t)s \in [0,1]$.
}

Although [7] considers only $m=2$, the proof easily extends to $m>2$.
Using this convergence result,
Theorem 5 also holds for smoothing splines:

{\bf Theorem 7.}{\em For a sequence of smoothing spline estimates of
$f$ as given by Thm.~6, Eq.~(4.2) holds 
provided that the smoothing parameters satisfy
$0< {\rm liminf}_n  \lambda_n^{1/2m}  n^{1/(2\ell+3)} \le $ 
${\rm limsup}_n  \lambda_n^{1/2m} n^{1/(2\ell+3)}<\infty $
and $\ell < 2m -5/2$.}


\sect{5. Data-based Pilot Estimators with Geometric Fidelity}
{\vskip-.1in} 

We consider two step estimators that begin by 
estimating $f^{(\ell)}$ and $f^{(\ell+1)}$ using an unconstrained estimate 
with $h_n\sim log^2(n)n^{1/(2\ell +3)}$. In the second step, we perform a
constrained fit, at some locations requiring 
$\widehat{f^{(\ell)}}$ to be monotone and in other regions 
 requiring $\widehat{f^{(\ell-1)}}$ to be monotone.
From the pilot estimate, we 
determine the number, $\kh$, and approximate locations 
of the inflection points.
At each empirical inflection point, $\hat{x}_j$, we define the $\alpha$
uncertainty interval by 
$[\hat{x}_j-z_{\alpha}\hat{\sigma}(\hat{x}_j),
\hat{x}_j+z_{\alpha}\hat{\sigma}(\hat{x}_j)]$,
where $\hat{\sigma}^2(\hat{x}_j)= 
\sigma^2\|\kappa^{(\ell)} \|^2  F'(s)/
|\widehat{f^{(\ell+1)}}(\hat{x}_j)|^2 n h^{2\ell +1}$
and $z_{\alpha}$ is the two sided $\alpha$-quantile for a normal distribution.

If an even number of uncertainty intervals overlap, we constrain the 
fit such that $\widehat{f^{(\ell)}}$ to be positive/negative in each interval.
If an odd number of uncertainty intervals overlap, we constrain the 
fit such that $\widehat{f^{(\ell+1)}}$ to be positive/negative 
in a subregion of the uncertainty interval which contains an even number 
of inflection points of  $\widehat{f^{(\ell+1)}}$. 
(The sign of $\widehat{f^{(\ell)}}$ or 
$\widehat{f^{(\ell+1)}}$ is chosen to match the outer region.)
Asymptotically, the uncertainty intervals do not overlap and we constrain the 
fit such that $\widehat{f^{(\ell+1)}}$ is positive/negative in each 
uncertainty interval.

{\bf Theorem 8.} {\em Consider a two stage estimator 
that with probability, $1 - {\cal O}(p_n)$, correctly
chooses a closed convex cone $V$, with $f\in V$,
in the first stage and then performs
a constrained regression as in (2.3) with $p=2$.  
For $f \in W_{m,2}$,
under the restrictions 
of Thm 4.4 of [9], 
the  estimate, $\fh$, converges as 
${\bf E}\| \fh -f\|^2_j \sim \alpha_j
\lambda^{(m-j)/m} \|f\|_m^2 + \beta_j\sigma^2 / (n \lambda^{2k+1\over 2m})$,
where $n$ is large enough that $\lambda_n  \|f\|^2_m  > p_n 
(1/n\sigma^2)\sum_i |f(t_i)|^2$ and 
$p_n n \lambda^{1\over 2m} \rightarrow 0$. 
}  

{\bfab Proof:} If the constraints are correct, Theorem 4 yields the 
asymptotic error bound [9]. We need to show that misspecified models do not
contribute significantly to the error.
If the  model is misspecified, then 
$\| \fh -f\|_V \le \lambda (\|f\|_m + \|\fh \|_m ) + 
(1/n\sigma^2)\sum_i |\fh(t_i)-f(t)-\epsilon_i |^2 +\epsilon_i^2$
$\le \lambda \|f\|_m + (1/n\sigma^2) \sum_i (y_i^2 +\epsilon_i^2)
\le \|f\|_V + 1.1 \chi_n^2(p_n)/n $.
The expected error is 
{\vskip-.07in} \noindent
$${\bf E}\| \fh -f\|^2_j \le
{\bf E}_{f \in V}\| \fh -f\|^2_j  + p_n 
{\bf E}_{f \notin V} 
\| \fh -f\|^2_j \ .$$  
{\vskip-.07in} \noindent
Note Theorem 4.4 of [9] applies to both pieces. 
Asymptotically as $p_n \rightarrow 0$,  
$\chi_n(p_n) \le 1.5n  + {\cal O}(p_n)$,
where $\chi_n^2(p_n)$ is defined by $\int^\infty_\chi dp_{\chi_n} =p_n $.  
\eopp


A similar result is given in [3] for the case of constrained least squares.
The trick of Theorem 8 is to constrain $\widehat{f^{(\ell+1)}}$ to be
positive (or negative) in  the uncertainty interval of the estimated 
inflection points
rather than constraining $\widehat{f^{(\ell)}}$ to have a single zero
around $\hat{x}_j$.

We recommend choosing the first stage halfwidth,
$h_n$ proportional to the halfwidth chosen by generalized crossvalidation
(GCV):
$h_n =\iota(n) h_{GCV}$ where $\iota(n) =  log^2(n)n^{1/(2\ell +3) -1/(2m+1)}$.
The second stage smoothing parameter, $\lambda_n$, is chosen to be the 
GCV value $\lambda_n =\lambda_{GCV}$. 
Other schemes [8] choose the final 
smoothing parameter to be the smallest value that yields only $k$
inflection points in an unconstrained fit. 
Since spurious inflection points asymptotically
occur only in a neighborhood of an actual inflection point,
these earlier schemes 
oversmooth away from the actual inflection points.
In contrast,
our second stage use the asymptotically optimal amount of smoothing
while preserving geometric fidelity.

\Acknowledgments
{Work funded by U.S. Dept.\ of Energy Grant DE-FG02-86ER53223.}
{\vskip-.1in}

\References

\ref{Aubin, J.-P.  and I.~Ekeland, {\em Applied Nonlinear Analysis},
John Wiley, New York 1984.}



\ref Gasser, Th. and M\"uller, H.,
Estimating 
functions and their derivatives by the kernel method,
{ Scand.~J.~of Stat.} {\bf 11} (1984), 171--185.

\ref Mammen, E., Nonparametric regression under qualitative
smoothness assumptions, Ann. Statist., {\bf 19} (1991), 741-759.

\ref Mammen, E., On  qualitative smoothness of kernel density estimates,
Univeristy of Heidelberg Report 614.

\ref Mammen, E., J.S.~Marron and N.J.~Fisher, 
Some asymptotics for multimodal tests based on kernel density estimates,
Prob. Th. Rel. Fields, {\bf 91} (1992), 115-132.



\ref{Michelli, C.~A., and F.~Utreras, 
Smoothing  and interpolation in a convex set  of  Hilbert space,
{SIAM J.~Stat.~Sci.~Comp.} {\bf 9} (1985),  728-746. }



\ref  Silverman, B.~W., Spline smoothing: the equivalent variable
kernel method,  Ann.~Stat. {\bf 12} (1984),  898-916.

\ref  Silverman, B.~W., Some properties of a test for
multimodality based on kernel density estimates, in ``Probability,
Statistics and Analysis'', J. F. C. Kingman and G. E. H. Reuter eds., pp.
248-259. Cambridge University Press, 1983.


\ref Utreras, F.,
Smoothing noisy data under monotonicity constraints - Existence,
characterization and convegence rates,
{Numerische Math.} {\bf 47} (1985),  611-625.

\ref  Wahba, G., {\it Spline Models for Observational Data}, SIAM,
Philadelphia, PA 1991.


\bigskip

\leftline{\it Kurt S.~Riedel}
\leftline{New York University, Courant Institute} 
\leftline{251 Mercer St., New York, NY 10012}
\leftline{\fab riedel@cims.nyu.edu}

\end
\end{document}

A similar theorem is given in [M] for constrained regression with no
penalty function. 
2) evaluate E[\# of 
$\ell$-inflection points in $\fh$]
using (5.?). If E[\# $\ell$-inflection points of $\fh$]
is greater than the number
of $\ell$-inflection points in our estimate plus $\alpha$,
increase $\lambda$ until Bias[\# of $\ell$-inflection points of $\fh$]

The negative polar of $V^{k,m }_{m,p} [\xbf]$ is
$V^{k,m }_{m,p} [\xbf]^- = -V^{k,m }_{m,p} [\xbf] \cap W_{m,p}^0$. 

The negative polar of $V^{k,0}_{1,p} [\xbf]$ is
$V^{k,0 }_{1,p} [\xbf]^- =$ closure in $W_{1,p}^*$ of
the $\{ g\ {\rm st}\  g''(t)f(t) \ge 0, f(1)g'(1) \le 0,  f(0)[g(0) -g'(0)] \le 0,\
{\rm for}\ f \in V^{k,0}_{1,p} [\xbf] $.

The negative polar of $V^{k,0}_{2,p} [\xbf]$ is
$V^{k,0 }_{2,p} [\xbf]^-$ = closure in $W_{2,p}^*$ of
the $\{ g\ {\rm st}\  g^{(iv)}(t)f(t) \le 0, g''(1)=0, g'(0) -g''(0) = 0,
f(1)g'''(1) \ge 0, f(0)[g(0)+g'''(0)] \le 0 \
{\rm for}\ f \in V^{k,0}_{2,p} [\xbf] $. 

The negative polar of $V^{k,m-1}_{m,p} [\xbf]$ is
$V^{k,m-1 }_{m,p} [\xbf]^-$ = closure in $W_{m,p}^*$ of
the $\{ g\ {\rm st}\  g^{(m+1)}(t)f^{(m-1)}(t) \ge
0, f^{(m-1)}(1)g^{(m)}(1) \le 0,  
f^{(m-1)}(0)[g^{(m-1)}(0) -g^{(m)}(0)] \le 0,\
{\rm for}\ f \in V^{k,m- 1}_{m,p} [\xbf] $.

The negative polar of $V^{k,m-2}_{m,p} [\xbf]$ is
$V^{k,m-2 }_{m,p} [\xbf]^- =$ closure in $W_{m,p}^*$ of
the $\{ g\ {\rm st}\  g^{(m+4)}(t)f^{(m-2)}(t) \ge
0, g^{(m+1)}(1) =0,  f^{(m-1)}(1)g^{(m)}(1) \le 0,  
f^{(m-1)}(0)[g^{(m-1)}(0) -g^{(m)}(0)] \le 0,\
g^{(iv)}(t)f(t) \le 0, g''(1)=0, g'(0) -g''(0) = 0,
f(1)g'''(1) \ge 0, f(0)[g(0)+g'''(1)] \le 0 \
{\rm for}\ f \in V^{k,m- 2}_{m,p} [\xbf] $. 

The negative polar of $V^{k,\ell}_{m,p} [\xbf]$ is
$V^{k,\ell }_{m,p} [\xbf]^- =$ closure in $W_{m,p}^*$ of
the $\{ g\ {\rm st}\  (-1)^{m-\ell}g^{(2m-\ell)}(t)f^{(\ell)}(t) \le 0, 
g^{(m-k-1)}(0) -(-1)^{k} g^{(m+k)}(0) = 0,
g^{(m+k)}(1)=0, k = 0 \ldots m - \ell -2,
(-1)^{\ell}f^{(\ell)}(1)g^{(2m-\ell-1)}(1) \ge 0, 
(-1)^{\ell}f^{(\ell)}(0)[g^{(\ell)}(0)- (-1)^{\ell}
g^{(2m-\ell-1)}(1) \ge 0, 
{\rm for}\ f \in V^{k,m- 2}_{m,p} [\xbf] $. 

For $V= V^{k,\ell}_{m,p}[\xbf]$, $N_V(f) =\{ g \in W_{m,p}^*
(-1)^{m-\ell}g^{(2m-\ell)}(t)f^{(\ell)
}(t) = 0, 
g^{(m+k)}(1)=0, g^{(m-k)}(0) -(-1)^{k+1} g^{(m+k)}(0) = 0,
(-1)^{\ell}f^{(\ell)}(1)g^{(2m-\ell-1)}(1) = 0, 
(-1)^{\ell}f^{(\ell)}(0)[g^{(\ell)}(0)- (-1)^{\ell}
g^{(2m-\ell-1)}(1) = 0.$

$
-V^{k,\ell }_{m,p} [\xbf] \cap W_{m,p}^0 | f^{(\ell )}
(t) g(t) \equiv 0\}$. Since $|f^{(\ell )} |>0$ implies $N_V(f)(t)=0$,
(2.4) follows. \eopp

\noindent
{\bf Lemma 1.} [1, p.171] {\em  Let $K$ be a closed convex cone in W, 
the normal cone of $K$ in $W^*$ at $f$, $N_K(f)$ satisfies 
$N_K(f)= K^- \cap \{ f \}^{\bot}$, where $K^-$ is the negative polar of $K$.
}

For $V= V^{k,\ell}_{m,p}[\xbf]$, $N_V(f) =\{ g \in
-V^{k,\ell }_{m,p} [\xbf] \cap W_{m,p}^0 | f^{(\ell )}
(t) g(t) \equiv 0\}$. Since $|f^{(\ell )} |>0$ implies $N_V(f)(t)=0$,
(2.4) follows. \eopp

\end{document}